\documentclass{appolb}
\usepackage{epsfig}

\begin{document}
\bibliographystyle{h-elsevier3} 
\title{Hydrodynamic flow from RHIC to LHC%
\thanks{Presented at \textit{Strangeness in Quark Matter 2011}, Sept. 18--24, Cracow, Poland.}%
}
\author{Piotr Bo\.zek
  \address{Institute of Nuclear Physics
    PAN,
    ul. Radzikowskiego 152, PL-31342 Krak\'ow, Poland} 
\address{Institute of Physics, Rzesz\'ow University, 
PL-35959 Rzesz\'ow, Poland}
}
\maketitle
\begin{abstract}
The hydrodynamic model for the expansion of the fireball 
in relativistic heavy-ion collisions is presented. 
Calculations using relativistic hydrodynamics of a fluid with small viscosity 
yield a satisfactory description of the experimental data on the particle 
spectra, the elliptic flow or the interferometry radii.
\end{abstract}
  
\section{Introduction}

The matter created in relativistic heavy-ion collisions is strongly interacting.
The dynamics of the system can be described as the relativistic expansion of 
a hot fluid \cite{Kolb:2003dz,Florkowski:2010zz}. Deviations from local 
equilibrium in the dynamical system lead to effects of viscosity in 
the  hydrodynamics \cite{Romatschke:2009im,Teaney:2009qa}. The collective
flow of the fluid is indirectly observed in the transverse momentum spectra 
of produced particles, in the azimuthally asymmetric directed, 
elliptic and triangular 
flows, and in the interferometry radii.  The experimental results 
for such soft observables  at the top RHIC and LHC
 energies can be quantitatively understood using hydrodynamic models 
\cite{Luzum:2008cw,Song:2010aq,Bozek:2009dw,Schenke:2010rr,Broniowski:2008vp,Pratt:2008qv}. 

Two conclusions on the nature of the hot and dense matter 
created can be deduced. First, the equation of state  at small baryon
 density has a crossover transition from the quark-gluon plasma to the
hadronic phase, in agreement with lattice QCD calculations \cite{Aoki:2006we}.
Second, the shear viscosity to entropy ratio is small $\eta/s < 0.2$, 
close to the estimates from strongly coupled theories \cite{Kovtun:2004de}.

\section{Hydrodynamics}

Second order viscous hydrodynamic equations 
\begin{equation}
\partial_\mu T^{\mu\nu}=0
\end{equation}
for the evolution in the proper time $\tau=\sqrt{t^2-z^2}$  of the  
energy momentum tensor are usually solved in $2+1$ \cite{Luzum:2008cw,Chaudhuri:2006jd,Song:2007fn,Dusling:2007gi,Bozek:2009dw,Schenke:2010rr}
(but also in   $3+1$ dimensions \cite{Schenke:2010rr,Bozek:2011ua}).
The energy momentum tensor contains corrections from 
 shear viscosity $\pi^{\mu\nu}$ and from  bulk viscosity $\Pi \Delta^{\mu\nu}$,
$\Delta^{\mu\nu}=g^{\mu\nu} -u^\mu u^\nu$, and $u^\mu$ is the fluid velocity.
The corrections are solutions of differential equations \cite{IS}
\begin{equation}
\Delta^{\mu \alpha} \Delta^{\nu \beta} u^\gamma \partial_\gamma \pi_{\alpha\beta}=\frac{2\eta \sigma^{\mu\nu}-\pi^{\mu\nu}}{\tau_{\pi}}-\frac{1}{2}\pi^{\mu\nu}\frac{\eta T}{\tau_\pi}\partial_\alpha\left(\frac{\tau_\pi u^\alpha}{\eta T}\right) 
\end{equation}
and
\begin{equation}
 u^\gamma \partial_\gamma \Pi=\frac{-\zeta \partial_\gamma u^\gamma-\Pi}{\tau_{\Pi}}-\frac{1}{2}\Pi\frac{\zeta T}{\tau_\Pi}\partial_\alpha\left(\frac{\tau_\Pi u^\alpha}{\zeta T}\right)  \ . \end{equation}
with  the shear  $\eta$ and bulk viscosity $\zeta$ coefficients,
 and the relaxation times $\tau_\pi$, $\tau_\Pi$.

We use a boost-invariant viscous hydrodynamic model 
with  parameters adjusted to RHIC data 
\cite{Bozek:2009dw} $\eta/s=0.08$ and $\zeta/s=0.04$ 
(bulk viscosity only in the hadronic phase). The initial profile 
of the  density in the transverse plane is taken from the Glauber model
\cite{Bozek:2010er,Bozek:2011wa,Bozek:2010bi}.
The emission of particles at the freeze-out temperature of $135$MeV 
is performed using the event generator THERMINATOR \cite{Kisiel:2005hn} 
including non-equilibrium corrections to the momentum distribution 
from shear and bulk viscosity.

\section{Results}

The directed flow in ultrarelativistic collisions measures the deflection 
of the fluid motion  from the beam axis. It can be a remnant of 
the initial flow or could result from  the early dynamics of a deformed 
fireball
 \cite{Snellings:1999bt,Csernai:1999nf}.
 In $3+1$-dimensional perfect fluid
 hydrodynamics, it can be generated in  the expansion of a  source
tilted away from the collision axis. The tilt 
of the source originates from the preferential emission in the 
forward (backward)
hemisphere from participant nucleons going in the forward (backward) direction
\cite{Bialas:2004su}.
\begin{figure}[t]
\begin{center}
\includegraphics[angle=0,width=0.59\textwidth]{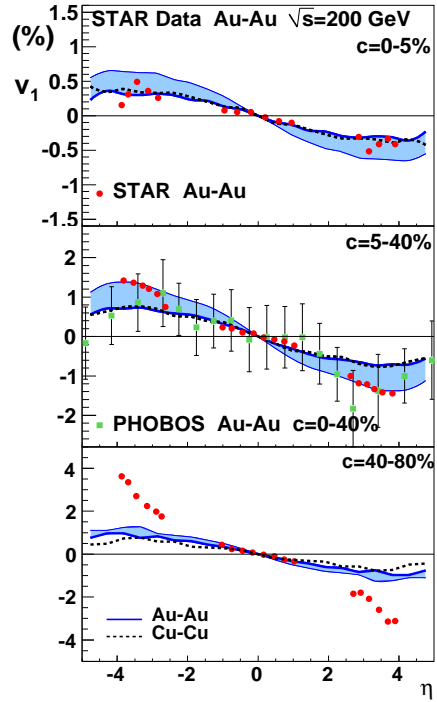}
\end{center}
\caption{\small  Directed flow in Au-Au  and Cu-Cu
   collisions (solid and dashed lines respectively) at
 different centralities from  hydrodynamic calculations,
 compared to the 
data of the  PHOBOS and STAR 
Collaborations
\cite{Back:2005pc,Abelev:2008jga}. The shaded band  between the thin 
and thick lines represents the effects of  the  uncertainty on the 
initial tilt of the source (from  \cite{Bozek:2010bi}).}
\label{fig:v1}
\end{figure}
Hydrodynamic results in Fig. \ref{fig:v1} show that the measured directed
 flow in central rapidities can be explained. As observed experimentally, 
the flow is similar in Au-Au and Cu-Cu collisions.

\begin{figure}[t]
\begin{center}
\includegraphics[angle=0,width=0.49\textwidth]{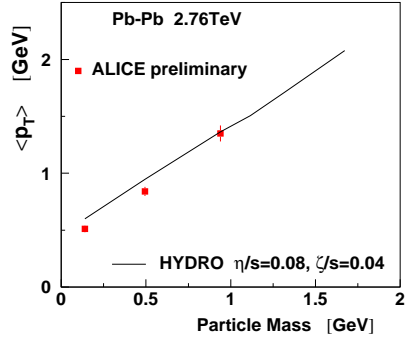}
\end{center}
\caption{\small Average $p_\perp$ of identified particles from
 viscous hydrodynamics \cite{Bozek:2011wa} compared to ALICE Collaboration
preliminary data  \cite{Floris:2011ru}.}
\label{fig:pt}
\end{figure}

The transverse momentum spectra of particles produced in Au-Au collisions
 at the top RHIC energies can be quantitatively described in hydrodynamic models
\cite{Florkowski:2010zz,Luzum:2008cw,Song:2010aq,Bozek:2009dw,Schenke:2010rr,Broniowski:2008vp,Chaudhuri:2006jd,Song:2007fn,Dusling:2007gi}.
The shape of the spectra results from a convolution of the collective flow of the fluid with the thermal emission at the freeze-out.
\begin{figure}[t]
\begin{center}
\includegraphics[angle=0,width=0.7\textwidth]{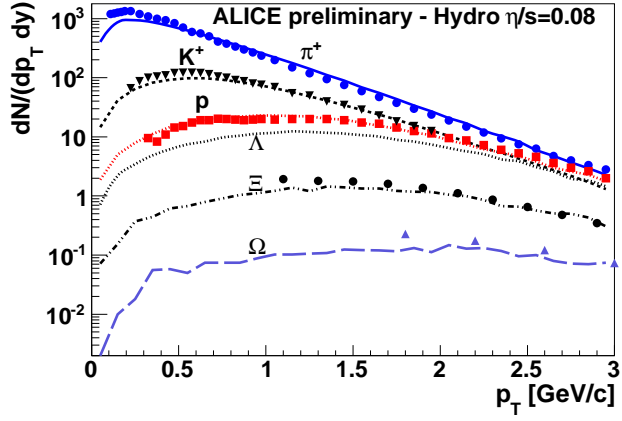}
\end{center}
\caption{\small Transverse momentum spectra of identified particles in Pb-Pb collisions. Preliminary data of the ALICE Collaboration
 ($\pi^+$, $K^+$, $p$) for centrality $0-5$\%  \cite{Floris:2011ru} and
 ($\Xi^-$, $\Omega^-$) for centrality $0-20$\% \cite{alicesqm} 
scaled by  $1.3$ to compare with
 viscous hydrodynamic results for the centrality $0-5$\% \cite{Bozek:2011wa}.}
\label{fig:spectra}
\end{figure}
In Pb-Pb collisions at $\sqrt{s}=2760$GeV, the transverse collective flow is 
stronger. One observes a significant shift of the spectra of protons towards higher $p_\perp$ (Fig \ref{fig:spectra}). For heavier particles (
$\Xi$ and $\Omega$) 
the flow predicted by hydrodynamic models is very strong, showing itself as an
 increase of the mean $p_\perp$ with the particle mass (Fig. \ref{fig:pt}).
The observed flow of heavy baryons is not as strong, which could indicate that 
multistrange baryons decouple earlier. Also their chemical decoupling temperature is higher than for protons \cite{alicesqm}.

\begin{figure}[t]
\begin{center}
\includegraphics[angle=0,width=0.49\textwidth]{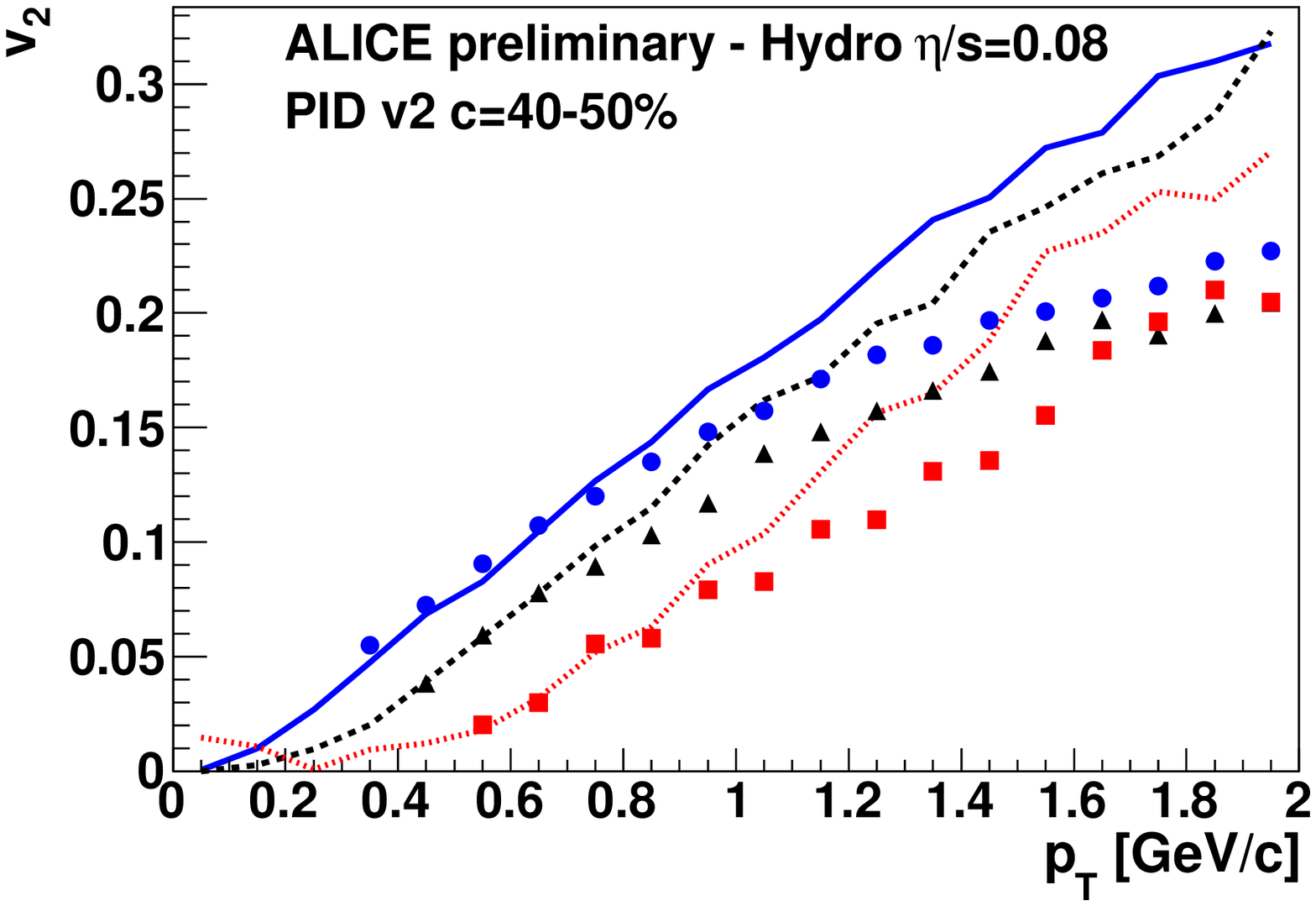}
\includegraphics[angle=0,width=0.49\textwidth]{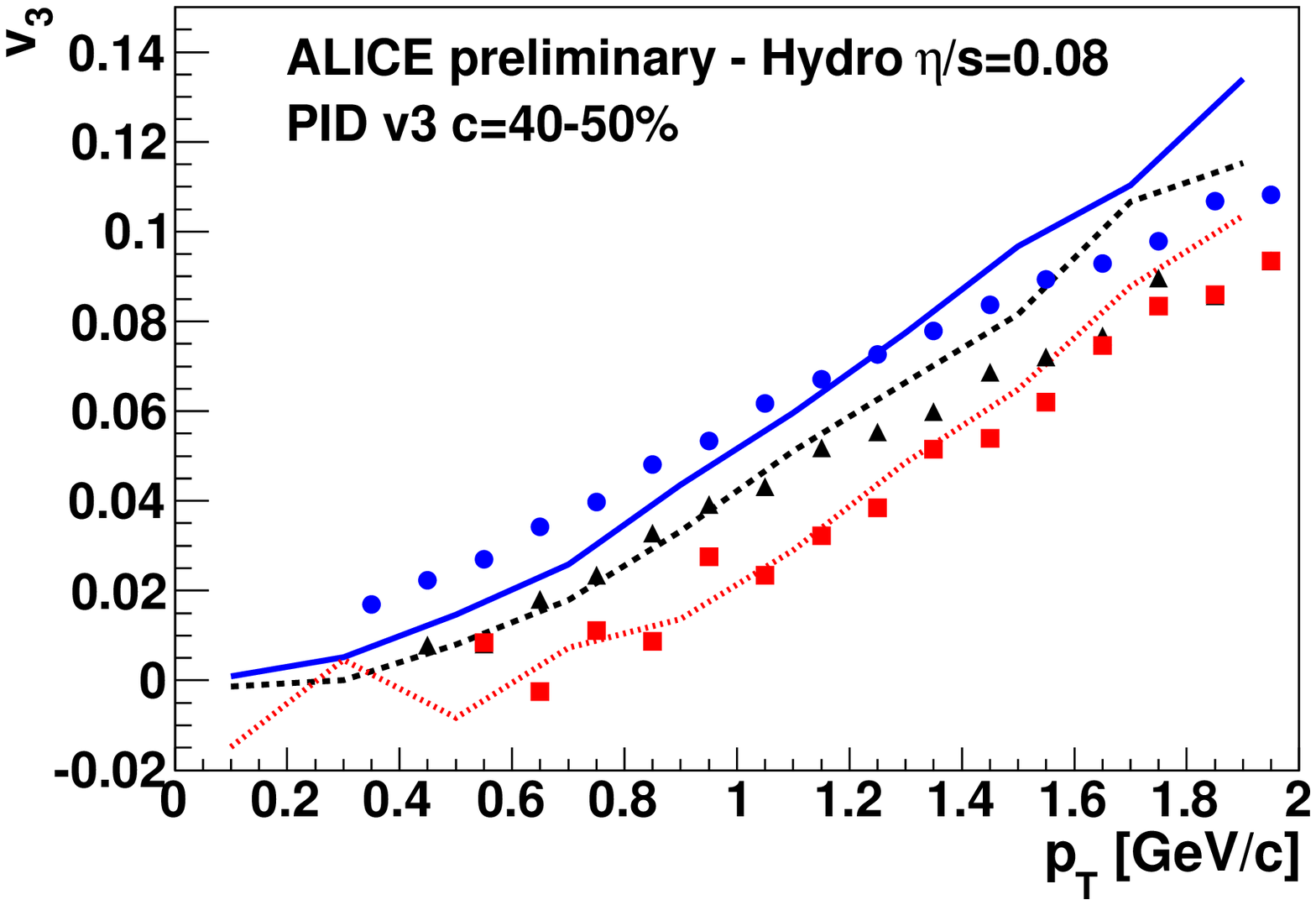}
\end{center}
\caption{\small (left pane) Elliptic flow of identified particles 
($\pi^+$, $K^+$, $p$ from top to bottom) from  
viscous hydrodynamics \cite{Bozek:2011wa} compared to ALICE Collaboration
preliminary data  \cite{Krzewicki:2011ee}. (right panel) Same 
but for the triangular flow.}
\label{fig:v2}
\end{figure}

The elliptic flow generated in the 
hydrodynamic expansion of a fluid with 
small viscosity is compatible with the observations \cite{Luzum:2008cw,Song:2010aq,Bozek:2009dw,Schenke:2010rr,Chaudhuri:2006jd,Song:2007fn,Dusling:2007gi,Bozek:2011wa}. The main uncertainty in the estimation of the viscosity 
coefficient from
phenomenological studies comes from the uncertainty in the
 value of the initial 
eccentricity of the fireball. The elliptic flow coefficient of charged particles 
as function of the transverse momentum is very similar at RHIC and at the LHC
 \cite{Krzewicki:2011ee}. The same is observed in hydrodynamic calculations, 
where the dependence of the elliptic flow $v_2(p_\perp)$ as function of energy saturates.
The elliptic flow of identified particles shows splitting for particles 
of different masses. At LHC the splitting is stronger as the transverse 
flow is more important (Fig. \ref{fig:v2}). The elliptic flow of protons,
 kaons and pions can be  well described in hybrid models connecting a 
hydrodynamic expansion stage with a hadronic cascade afterburner
 \cite{Heinz:2011kt}. 

The triangular flow reflects shape fluctuations in the source
 \cite{Alver:2010gr,Petersen:2010cw,Alver:2010dn,Schenke:2010rr,Qiu:2011fi}. 
The collective expansion of the fireball with a triangular deformation 
is very sensitive to the value of the shear viscosity.
A common description of the elliptic and triangular flows 
could constraint the initial fluctuations and the value of the parameters.
In Fig. \ref{fig:v2} is shown the triangular flow of identified 
particles at the LHC. In our calculation, 
the triangular deformation is added to the optical Glauber model
 density following the prediction of the 
 Glauber Monte-Carlo model. We observe a similar splitting in the
value of $v_3$ between particles of different mass as for the elliptic 
flow coefficient. However, the same calculations that describes the elliptic 
flow of identified particles cannot reproduce the values for the
 triangular flow.

\begin{figure}[t]
\begin{center}
\includegraphics[angle=0,width=0.44\textwidth]{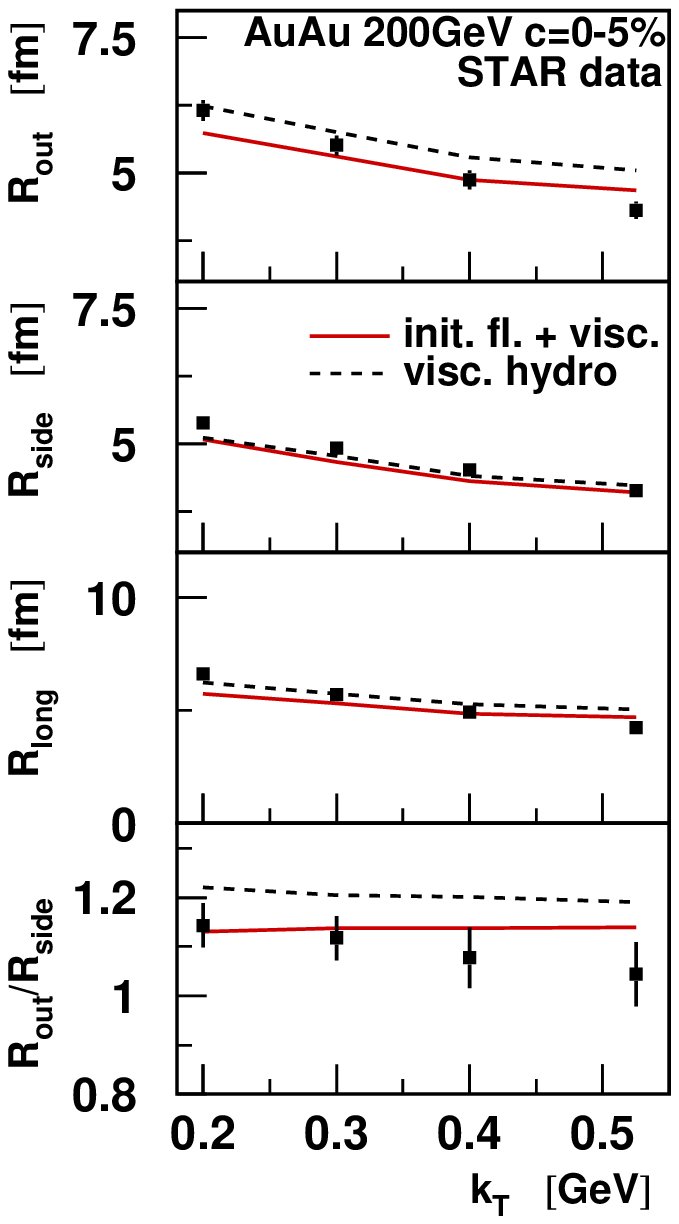}
\includegraphics[angle=0,width=0.45\textwidth]{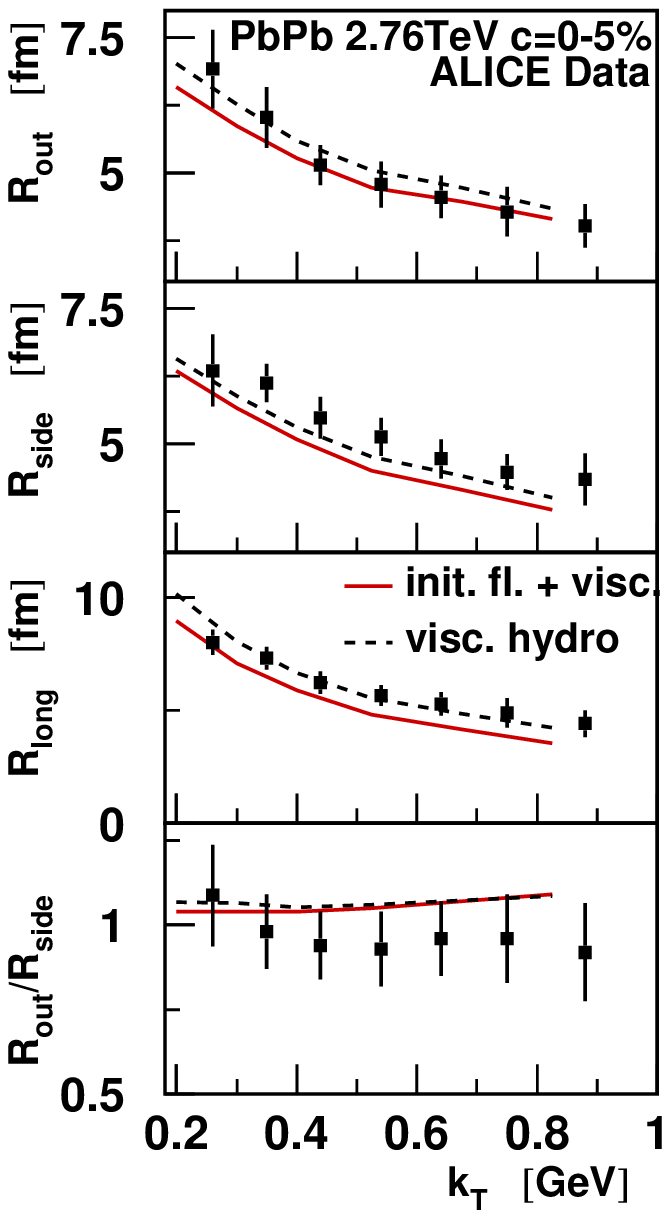}
\end{center}
\caption{\small Interferometry radii for Au-Au collisions 
at $\sqrt{s}=200$GeV \cite{Adams:2004yc} (left panel) and Pb-Pb collisions at $\sqrt{s}=2760$GeV  \cite{Aamodt:2011mr}
 (right panel) compared to viscous hydrodynamic calculations \cite{Bozek:2010er}}
\label{fig:hbt}
\end{figure}

The identical particle interferometry is an important tool for measuring
 the size and the life-time of the interacting system 
\cite{Florkowski:2010zz,Lisa:2005dd}. Hydrodynamic calculations with a hard equation of state yield reasonable values of the extracted interferometry radii
(Fig. \ref{fig:hbt}) at RHIC energies 
\cite{Broniowski:2008vp,Pratt:2008qv,Karpenko:2009wf,Bozek:2010er,Kisiel:2008ws,Bozek:2010wt}. At the LHC, the hydrodynamic 
transverse flow is stronger which gives an even better agreement of the interferometry radii with the data. We notice that nonzero viscosity or the presence of the  pre-equilibrium flow  improve
 the agreement for the $R_{out}/R_{side}$ ratio.

The high  multiplicity of particles created in proton-proton collisions 
at LHC energies would indicate that some degree of collectivity could 
be achieved in the interaction region. Effects of the 
collective expansion of the matter
 in such collisions  have been 
estimated 
\cite{Luzum:2009sb,Bozek:2009dt,CasalderreySolana:2009uk,Avsar:2010rf}. 
The observation of the ridge in the two-particle correlations in
high multiplicity events in proton-proton collisions by the
 CMS Collaboration \cite{Khachatryan:2010gv} can be 
interpreted as the existence of the
 elliptic flow \cite{Bozek:2010pb,Werner:2010ss}. However, it is difficult to
separate it from important non-flow effects. The presence of a significant 
 collective flow could be 
observed  through  interferometry methods applied to proton-proton collisions
\cite{Aamodt:2011kd,Bozek:2009dt,Kisiel:2010xy,Werner:2011fd}.

\section{Conclusions}

The hydrodynamic expansion of a hot and dense  fluid 
represents  a realistic model of the 
dynamics of the fireball in relativistic heavy-ion collisions. 
At RHIC energies the viscous hydrodynamic model can describe 
the particle spectra, the elliptic flow of charged and identified particles, 
the interferometry radii, and the triangular flow. Different 
implementations of the hydrodynamic model in the literature, especially with respect to the initial conditions and the 
final hadron rescattering give a better agreement for some of the soft
above mentioned 
observables than for  other.

Using the hydrodynamic model for Pb-Pb collisions at the LHC, one finds a
satisfactory description of most of the available data at soft transverse
 momenta. The collective transverse flow is stronger and as a consequence
 the mean $p_\perp$ of produced
 particles and the integrated elliptic flow increases as compared to RHIC. 
The interferometry radii show a dependence on the momentum of the  pion pair 
characteristic for the collective flow. The elliptic flow of charged 
particles as function of $p_\perp$ is hydrodynamically saturated
 and does not change significantly when $\sqrt{s}$ increases  from $200$ 
to $2760$GeV. The strong collective component in the spectra is
visible for pions, kaons and protons, but is less so for multistrange 
baryons. This may indicate an earlier decoupling of heavy strange particles.
The strong  flow explains the particle  mass splitting for the
elliptic flow but cannot explain the one for the triangular flow.
The appearance of the collective expansion in high multiplicity proton-proton
 collisions has been suggested, but it is difficult to 
evidence it experimentally.

\section{Acknowledgments}
This work was supported in part by the MNiSW grant No. N N202 263438. 

\bibliography{../hydr}
\end{document}